\documentclass[runningheads]{svmult}
\usepackage{epsfig} 

\usepackage{makeidx}   % allows index generation
\usepackage{graphicx}  % standard LaTeX graphics tool
\usepackage{subeqnar}  % subnumbers individual equations
\usepackage{multicol}  % used for the two-column index
\usepackage{physprbb}  % modified textarea for proceedings,
\makeindex             % used for the subject index
\def\lapprox{\mathrel{\mathop
  {\hbox{\lower0.5ex\hbox{$\sim$}\kern-0.8em\lower-0.7ex\hbox{$<$}}}}}
\def\gapprox{\mathrel{\mathop
  {\hbox{\lower0.5ex\hbox{$\sim$}\kern-0.8em\lower-0.7ex\hbox{$>$}}}}}

\begin{document}
\title*{$\alpha$: a constant that is not a constant?}
\toctitle{$\alpha$:a constant that is not a constant?}
% allows explicit linebreak for the table of content
%
%
\titlerunning{$\alpha$:a constant that is not a constant?}
% allows abbreviation of title, if the full title is too long
% to fit in the running head
%
\author{G. Fiorentini
\and B. Ricci}
\authorrunning{G. Fiorentini et al.}
% if there are more than two authors,
% please abbreviate author list for running head
%
%
\institute{Dipartimento di Fisica, Universit\'{a} di Ferrara 
and Istituto Nazionale di Fisica Nucleare, 
Sezione di Ferrara,\\
I-44100 Ferrara, Italy}

\maketitle              % typesets the title of the contribution

\begin{abstract}
We review the observational information on the constancy of the fine
structure constant $\alpha$. We find that small improvements on the 
measurement of $^{187}Re$ lifetime can provide significant 
progress in exploring the range of variability  suggested by QSO data.
We also discuss  the effects of a time varying $\alpha$ on 
stellar structure and evolution.
We find that radioactive dating of ancient stars can offer a new
observational window. 
\end{abstract}

\section{Introduction}

The possibility that some of 
the ``fundamental constants'' may depend 
on time was first discussed by Dirac \cite{Dirac1}.
He remarked  that the huge  ratio 
of electric to gravitational forces between a proton and an  electron,
about $10^{39}$, was of the same order of magnitude as 
the age of the universe  in units provided by the atomic 
constants, $e^2/m_e c^3$. 
If this coincidence is not casual, 
then one must have varying constants, their values 
changing as the age of the universe changes:
\emph{``This suggests that the above mentioned large numbers are to be 
regarded not as constants, but as simple functions of our 
present epoch, expressed in atomic units .... 
In this way we avoid the need of a theory to determine numbers of 
the order $10^{39}$.''}
The approach of Dirac to what is now called the  hierarchy problem opened
a rich field of investigation. The variability of fundamental 
constants was analysed by Gamow, Dyson and others and then it
was forgotten for a while.

Interest in this topic has been revived in the context of string 
theories, where all the coupling constants and parameters, 
except the string tension, are actually derived quantities, 
which are determined by the vacuum expectation values of the 
dilaton and moduli. Since all these fields evolve on cosmological scales
the time variation of the constants of nature during the evolution 
of the universe arises as a natural possibility, 
see e.g. \cite{DP,Witten}.

On the observational side, Webb et al. \cite{Webb} have
presented evidence for a cosmological evolution
of the fine structure constant $\alpha=e^2/\hbar c$.
 The absorption spectra of diffuse clouds 
illuminated by quasars suggest that   ten billion years ago  $\alpha$ was slightly 
smaller, by about ten part per million.
Of course this indication, if confirmed, would have enormous importance. 

This short review attempts to provide an answer to some natural 
questions following the  claim of ref.\cite{Webb}:
 
i)What are the observational constraints on the variability of $\alpha$
and how do they compare with the result of ref. \cite{Webb}?

ii)What  are the prospects for improvements?

iii)What  are the effects of a time  varying $\alpha$ on stellar structure and evolution?

%%%%%%%%%%%%%%%%%%%%%%%%%%%%%%%%

\begin{figure}[htb]
\begin{center}
\includegraphics[width=0.5\textwidth, angle=270]{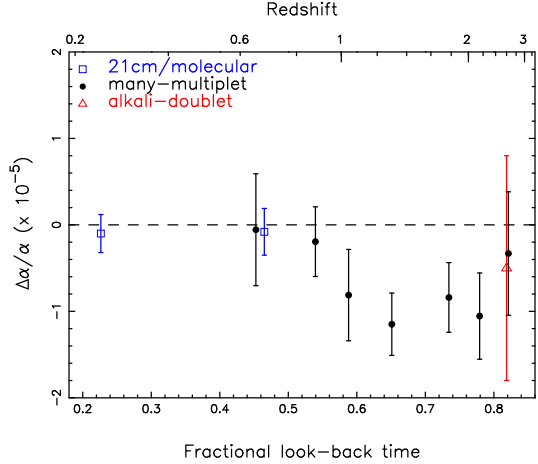}
\end{center}
\caption[]{
$\Delta \alpha/\alpha$ vs. fractional look-back time to the Big Bang,
from \cite{Webb}. 
%The conversion between redshift and look-back time assumes H$_0 = 68$
%km/s/Mpc, $(\Omega_{\rm M}, \Omega_{\rm \Lambda}) = (0.3, 0.7)$, so
%that the age of the universe is 13.9 Gyr.  72 quasar absorption
%systems contribute to this binned-data plot, 
%The hollow squares
%correspond to two HI 21cm and molecular absorption systems
%\cite{murphy_rad}.  Those points assume no change in $g_p$, so should
%be interpreted with caution.  The 7 solid circles are binned results
%for 49 quasar absorption systems.  The lower redshift points (below $z
%\approx 1.6$) are based on (MgII/FeII) and the higher redshift points
%on (ZnII, CrII, NiII, AlIII, AlII, SiII) \cite{murphy_mm}. 28 of these
%49 systems correspond to the sample used in \cite{webb}.  The hollow
%triangle represents the average over 21 quasar SiIV absorption
%doublets using the alkali doublet method \cite{murphy_si}.
}
\label{Figqso}
\end{figure}

%%%%%%%%%%%%%%%%%%%%%%%%%%%%%%%%%%%%

\section {What do quasars tell us?}

The measurement of  the spectra of distant quasars 
as a mean to study possible  variations of $\alpha$ was 
first suggested  by Savedoff \cite{Savedoff}.
Narrow lines in quasar spectra are produced by 
absorption of radiation in intervening clouds of gases.  
Essentially one needs to identify two (sets of) lines, 
which depend differently on $\alpha$, so as to 
extract the value of the redshift factor $z$ together with 
the value of $\alpha$ at that epoch. 
The fine structure doublets of ``alkali atoms''--  a term used
to denote   atoms and atomic ions 
with just one electron in the outer shell  -- are well suited 
for this study.

This method has been  used by several authors and it has
been  recently applied to a selection 
of high resolution observations, see \cite{VPI}. No indication of a 
variable $\alpha$ has been found and the constraint
$\Delta \alpha / \alpha 
= (-4.6 \pm 4.3[\mbox{stat}] \pm 1.4[\mbox{sys}] ) 10^{-5}$
has been obtained \cite{VPI} on the possible deviation at 
$z=2\div4$ from the present ($z=0$)value.

On the other hand, Webb et al. \cite{Webb} have used a ``many multiplet'' method, 
where $\alpha$ is estimated from comparison of 
the lines of \emph{different} atomic species,
so as to obtain a sensitivity gain. The data are summarized in Fig. 
\ref{Figqso}.
In this way they claim to have found  a deviation from the present $\alpha$ 
value over the redshift range  $z=1\div3$ :
\begin{equation}
\Delta \alpha / \alpha = (-0.72 \pm 0.18 ) \cdot 10^{-5} 
\end{equation}

This result has been criticized in ref. \cite{VPI} on the grounds 
that some systematic effect could mimic the variation of 
$\alpha$. For example, 
the lines of the two atomic species considered 
in \cite{Webb} are situated in different regions, so that calibration
errors could simulate the effect of $\alpha$ variation.
In contrast, the method based on the fine 
splitting of a line of the same species 
is not affected by these uncertainty sources.

%%%%%%%%%%%%%%%%%%%%%%%%%%%%%%%%%%%%%%%%%%%%%%%%%%%%%%%%%
\begin{table}[htb]
\caption{Summary on the variation of $\alpha$ }
\begin{center}
\renewcommand{\arraystretch}{1.4}
\setlength\tabcolsep{9pt}
%\begin{tabular}{@{}llp{1.8cm}l}
\begin{tabular}{p{2.2cm}lllll}
\hline\noalign{\smallskip}
Source &$\Delta \alpha /\alpha$ & Look back  & $z^*$  & $\dot\alpha/\alpha$ & ref.\\
       &                        & time (Gyr)           &    &     (yr$^{-1}$)&        \\
\noalign{\smallskip}
\hline
\noalign{\smallskip}
Laboratory & $ \leq 1.6 \cdot 10^{-14}$ & $ 4 \cdot 10^{-10}$ & 0     & $ \leq 4 \cdot 10^{-14}$ &\cite{lab}\\
Oklo       & $ \leq 1 \cdot 10^{-7}$    &  1.8       &  $\simeq 0.1 $ & $ \leq 6 \cdot 10^{-17}$ &\cite{oklo3}\\
Meteorites & $\leq 4  \cdot 10^{-6}$   &   4.5       &  $\simeq 0.4 $ & $ \leq 2 \cdot 10^{-15}$ & --\\
$^{12}$C & $\leq  10^{-2}$   &   $\simeq 10$       &  $\simeq 1.5 $ & $ \leq 10^{-12}$ &--\\
stellar dating      & $ \leq 10^{-3}$      & $\simeq 10 $& $\simeq 1.5$        & $  \leq 10^{-13}$    & --\\
QSO(doublet)  & $ \leq 10^{-4}$      & $\simeq 11 - 13 $&    2--4        & $  \leq 10^{-14}$    &\cite{VPI}\\
\textbf{QSO(multiplet)}        & $\mathbf{ + 1 \cdot 10^{-5}}$      & $\mathbf{\simeq 8 - 12 }$& \textbf{1--3}        & $\mathbf{ + 1 \cdot 10^{-15}  }$  &\textbf{\cite{Webb}}\\
CMB        & $ \leq 5 \cdot 10^{-2}$   & $\simeq 14 $&  $\simeq 10^3 $& $ \leq 3 \cdot 10^{-12}$ &\cite{Avelino}\\
BBN        & $ \leq 1 \cdot 10^{-2}$   & $\simeq 14 $ & $\simeq 10^9 $& $ \leq 7 \cdot 10^{-13}$ &\cite{Avelino}\\
\noalign{\smallskip}
\hline
\noalign{\smallskip}
\end{tabular}
\end{center}
$^*$ The red shift -- time connection is estimated for 
$H_o=68$ Km/s/Mpc, $\Omega_M=0.3$ and $\Omega_\Lambda =0.7$ ($t_u\simeq14$ Gyr). \\
%$^{\mathrm b}$ One second of arc.
\label{Tabalfa}
\end{table} 
%%%%%%%%%%%%%%%%%%%%%%%%%%%%%%%%%%%%%%%%%%%%%%%%%

\section{Quasars and the rest of the world}

The available information on the variability of  $\alpha$ is summarized in Table \ref{Tabalfa}.
Measurement in the laboratory
 are sensitive to extremely tiny variations $\Delta \alpha/\alpha \simeq 10^{-14}$,
 however on a   time scale of just a few months.
Essentially, one is comparing two clocks
(a Hg$^+$ atomic clock and a Hydrogen maser), with frequencies which depend 
differently on $\alpha$ \cite{lab}. 
Experiments with cold atoms will provide a significant sensitivity gain.
In fact, the ultimate limit for frequency 
measurement is observation time. Cold atoms in the laboratory  
fall due to gravity, whereas atoms in free fall do not fall at all, 
so let's go to space.
 This is the idea of  an extremely interesting project on the International 
Space Station, which is expected to  explore changes  
of $\alpha$ to the level  
  $\Delta\alpha/\alpha \simeq 10^{-16}$ \cite{spazio}.

The  physics of the fission reactor  which nature 
operated at Oklo about two billion 
years ago provides a very important 
 constraint. The footprints of  natural 
fission arise from
the abundances of rare earth isotopes at the Oklo site, which look similar to 
those produced by fission {\it today}.
These isotopic abundances are related  to large capture cross section of thermal neutrons, 
which correspond to
nuclear resonances at about the thermal energies. The similarity of the abundances means 
thus that, in two billion years,  nuclear energy levels has not varied by more than  
$kT\simeq 0.1$ eV, a very small range in comparison with the nuclear physics scale. The Coulomb contribution to the difference of nuclear energy levels.
$E_{Cou} \simeq  \alpha/r_{nuc}$, is
thus strongly fixed, corresponding to $|\Delta \alpha/\alpha| \lapprox 10^{-7}$ (barring 
from accidental cancellation 
due to variations of other fundamental parameters).  This was first pointed out in \cite{oklo1}
 and then discussed with much greater detail in \cite{oklo2,oklo3}. 

The Oklo bound arises, essentially, from the fact that the  Q value of a nuclear
 reaction contains an electromagnetic contribution which is sensitive to changes 
of $\alpha$. The constancy of Q, within a small scale of order $kT$, follows from 
the observation that reaction rates are the same, now and at the Oklo time.  
Conceptually, one is again  comparing two (nuclear) clocks
operating at different times.  

A similar argument can be applied to radioactive dating methods.  
The point is  that  the  lifetimes $\tau$ of radioactive nuclei  
depend on the Q-value of the decay. In addition,  $\alpha$-decay rates 
have an exponential dependence on $\alpha$, corresponding to the 
exponentially small tunnel probability.
The most sensitive process is $^{187}Re \rightarrow ^{187}Os +e + \bar\nu$
due to a very small Q-value:  
$\Delta \tau/ \tau \simeq 1.8 \cdot 10^4 \Delta \alpha /\alpha$ \cite{Dyson}, see Table \ref{Tabs}.
The laboratory measurement  
$\tau_{1/2}(lab)= (42.3\pm0.7)$Gyr (68\%C.L.) \cite{Lindner}
can thus be compared \cite{Allegre} with the value inferred from Re/Os 
measurement in ancient meteorites
$\tau_{1/2}(met)= (41.6\pm0.42)$ Gyr  \cite{Smoliar},
 dated by means of different radioactive methods 
(e.g. U/Th method, which is much less weakly 
affected by variation of $\alpha$).
 The agreement within errors  
(again apart for accidental cancellations)
provides a significant constraint, 
$ \overline{\Delta\alpha}/\alpha=(1\pm1)10^{-6}$,
where the bar denotes an average over the meteorite lifetime.
 It is not as strong as the Oklo bound, however it explores 
earlier times. Furthermore, 
two independent constraints 
(Oklo and meteorites) are important if one consider the possibility of simultaneous variations 
of several fundamental parameters.

Cosmic microwave background (CMB) yields information on the variability of $\alpha$ 
at even earlier times, since decoupling between radiation and matter occurs at 
the recombination epoch, the time when temperature is so low that atoms can be
stable. This clearly depends on the atomic binding energy and thus on $\alpha$ \cite{Kolb}. 
A change of  $\alpha$ also affects  the Big-Bang Nucleosynthesis  (BBN) mainly through the 
neutron-proton mass difference, which fixes the neutron density at the weak interactions 
freeze-out and consequently the primordial $^4He$ abundance.
According to most recent analysis, both  CMB and BBN are consistent with a a constant 
$\alpha$ and constrain $\Delta \alpha /\alpha$ to the per cent level, 
at $z \simeq 10^3$ and $10^9$ respectively \cite{Avelino}. 

For comparing the different information one has to make assumptions about  
the time  evolution of
$\alpha$. 
The simplest hypothesis is  a linear time dependence, which is used in the fifth column
of  Table \ref{Tabalfa}. In this case the QSO positive result only 
conflicts with the Oklo bound,
which provides the most strict constraint. 
It is interesting to observe that the meteorites give a bound close to the QSO signal. 
Improvements in the laboratory measurements of the $^{187}Re$ lifetime would thus be relevant.
The planned atomic physics experiment on the ISS  should  be capable of exploring 
the region suggested by QSO.

Let us remark that the QSO-Oklo apparent conflict 
can be avoided if  linearity does 
not hold. In addition $\alpha$ could depend on place.
Also one has  to remind that $\alpha$ is not alone;
its evolution has to be accompanied by the evolution
 of other coupling constants, otherwise unification of interactions  at the present 
epoch is just occasional. Actually, unification requires that a change of $\alpha$ 
is accompanied  by a much stronger  change in strong interaction parameters, 
see e.g. \cite{Lang}. 
A change of $\alpha$ corresponds to  a change of the QCD scale parameter 
$\Delta \Lambda_{QCD}/ \Lambda_{QCD} \simeq 40 \Delta\alpha/\alpha $.
This has important consequences, since nucleon and pion masses scale as 
$M_n \propto \Lambda_{QCD}$ and $M_{\pi}\propto  \sqrt{\Lambda_{QCD}}$.
Nuclear radii, which depend on the range of the nuclear force, are
thus also sensitive to a change of $\Lambda_{QCD}$.

In this situation, the analysis becomes much more complex, 
see e.g. \cite{Lang,Fritsch,Olive,Flam}
and a complete discussion has not yet been performed. Generally, one can remark the following points: 

-Information on the change of $\alpha$ give also information on the couplings of other interactions.

-Several different sources of information are needed, for disentangling the contributions of different effects.

For these reasons, improvements of the various methods, which explore different space-time 
regions and receive contributions from different interactions,  are important 
so as  to confirm or constrain a possible variation
 of  fundamental ``constants".

\begin{table}[htb]
\caption{  $\alpha$--dependence of nuclear halflives, from \cite{Dyson}}
\begin{center}
\renewcommand{\arraystretch}{1.4}
\setlength\tabcolsep{15pt}
%\begin{tabular}{@{}llp{1.8cm}l}
\begin{tabular}{llll}
\hline\noalign{\smallskip}
Nucleus & Decay  & $\tau_{1/2}$ [yr] & d(ln$\tau$)/d(ln$\alpha$)\\
\noalign{\smallskip}
\hline
\noalign{\smallskip}
$^{238}$U & $\alpha$ & $2 \cdot 10^9$ & $\simeq -500$ \\
$^{40}$K & EC & $1.3 \cdot 10^9$ & $\simeq +30$ \\
$^{187}$Re & $\beta$ & $4 \cdot 10^{10}$ & $\simeq +18000 $ \\
\noalign{\smallskip}
\hline
\noalign{\smallskip}
\end{tabular}
\end{center}
\label{Tabs}
\end{table}

%%%%%%%%%%%%%%%%%%%%%%%%%%%%%%%%%%%%%%%%%%%%%%%%%

\section{Stars and $\alpha$}

Stars are  a useful laboratory for studying fundamental physics. 
In principle, at least,  a change of $\alpha$ over very long times can 
affect stellar structure and evolution. Clearly, a change of  $\alpha$
 will affect  nuclear fusion rates and opacity. For the   former,  
one has  an exponential effect  in the tunnelling probability, 
whereas opacity scales with a small power of $\alpha$ 
(e.g.  $\kappa \propto \alpha^2$for Thomson scattering ). 
In this spirit we shall briefly discuss a few relevant points:

\subsection{The Sun}

\begin{figure}[htb]
\begin{center}
\includegraphics[width=0.5\textwidth]{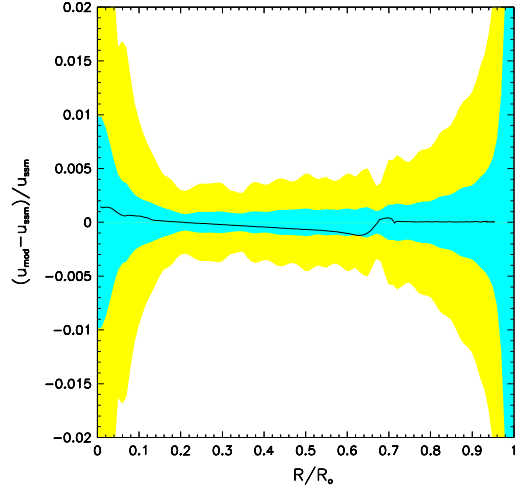}
\end{center}
\caption[]{Relative difference (model-SSM)/SSM of (isothermal) sound speed squared 
as a function of the radial coordinate for a time variation of $\alpha$ 
such that  the difference between sun
formation and present is $\Delta\alpha/\alpha = 10^{-2}$.
The $1\sigma$ ($3\sigma$) helioseismic uncertainty corresponds to the dark (light) area,
from \cite{eliosnoi}.
}
\label{Figelio}
\end{figure}

 Helioseismology has provided us with a  detailed knowledge of the  
present solar interior, well in agreement with Standard Solar Model (SSM) 
calculations which have been performed by 
assuming that  $\alpha$  has been constant. 
Can a  time dependent $\alpha$ spoil 
this agreement? We have constructed solar models with a linearly 
 time dependent  $\alpha$ such that  the difference between solar 
formation and present is $\Delta\alpha/\alpha = 10^{-2}$.
Note that this is a much larger variation than that 
implied by Oklo and/or meteorite constraints. 
The tiny difference in sound speed with respect to the 
SSM is well within the errors of helioseismic 
determination, see Fig. \ref{Figelio}.

%%%%%%%%%%%%%%%%%%%%%%%

%\begin{figure}
%\begin{center}
%\epsfig{file=fig_elio.ps,height=17cm}
%\end{center}
%\caption{The normalized response functions of SK and SNO to
%$^{8}B$ neutrinos, for representative values of detector
%thresholds. See text for details.}
%\label{f20}
%\end{figure}

%%%%%%%%%%%%%%%%%%%%%%%%

\subsection{Globular clusters}
\begin{figure}[htb]
\begin{center}
\includegraphics[width=0.5\textwidth]{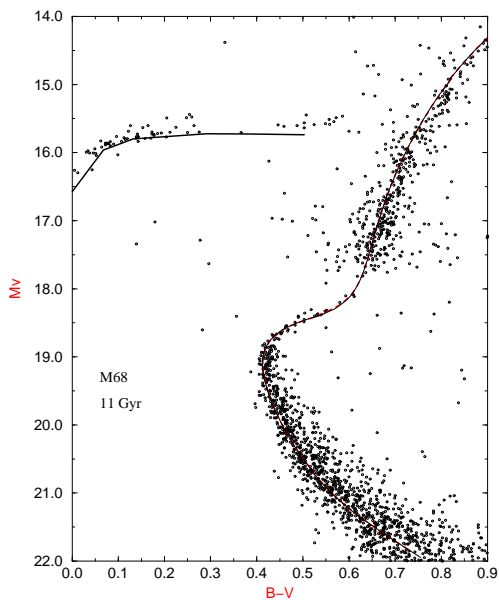}
\end{center}
\caption[]{Isochrones calculated  with constant $\alpha$ (solid line) and 
with a 
time dependent  $\alpha$ such that  the difference between cluster 
formation and present is $\Delta\alpha/\alpha = 10^{-2}$ (red dashed line).
The two curves look almost undistinguishable.}
\label{Figcluster}
\end{figure}

As well known, the evolution of these  systems  provides 
 a powerful method for determining  the age of the Galaxy, see. e.g. \cite{scilla}. 
Is this 
dating method affected by a time variation of $\alpha$?
Again the answer is negative, even for variation at the 
percent level over the Galaxy age, see Fig. \ref{Figcluster},
where we present  isochrones at $t=11$ Gyr calculated 
for the M68 cluster, as an illustrative example.

%%%%%%%%%%%%%%%%%%%%%%%
%%%%%%%%%%%%%%%%%%%%%%%%

\subsection{Stellar nucleosynthesis of $^{12}C$}

Our very existence relies on a nuclear accident, 
i.e. a suitably $^{12}C$ excited level which allows  
the carbon synthesis  by means of  
$\alpha+ \alpha+\alpha \rightarrow ^{12}C^* \rightarrow ^{12}C+\gamma $. 
 Carbon synthesis occurs at $kT \simeq 10$KeV and the resonance
 position is measured  at $(m_{12}^*-3 m_4)c^2=379.5$Kev. 
The observation of carbon  in ancient stars
implies that  some 10 Gyr ago the resonance energy was  
the same, within $kT$. Thus the Coulomb contribution  to the energy difference levels
(about $\alpha/r_{nuc}$) has not changed  by more than $kT$, which implies 
$\Delta\alpha/\alpha \le 10^{-2}$. Essentially, this is the same 
argument as for the Oklo reactor, however the bound is weaker since $kT$ is larger.

\subsection{Radioactive dating  of ancient stars}

In the last few years, radioactive dating has been extended beyond 
the solar system, see e.g.  \cite{Truran}. 
Thorium  dating of field halo stars and globular cluster stars yields
 ages on the order of $(15 \pm 4)$ Gyr, in agreement with the value derived
from globular cluster evolution.
 Furthermore,  recently the age  of an old star   
($\tau \simeq 12$ Gyr) has been determined by means of \emph{both}   Th and U dating \cite{Cayrel} 
so that two clocks are available!  The two methods are in agreement within  errors 
of about 3 Gyr, under the assumption  that  nuclear lifetimes have remained constant. 
By exploiting the different $\alpha$ dependence of the decay rates, 
(from \cite{Dyson} one derives 
$\frac{d(ln\tau)}{d(ln \alpha)}= -450$ and $-470$ for $^{238}U$ and $^{232}Th$ respectively) 
the coherence of 
results implies that  $\alpha$ has remained constant to the level 
$\Delta \alpha /\alpha \simeq 2.5 \cdot 10^{-2}$ on a 12 Gyr scale. 
There is a substantial cancellation of the $\alpha$ varying effect due to the similar
$\alpha$--dependence of the two nuclear clocks.
One can achieve a stronger constraint by comparing the Uranium clock with
the dating provided by globular cluster evolution
(which is not affected by $\alpha$ changes). In this case one has 
$\Delta \alpha /\alpha \lapprox 10^{-3}$.

The measurement of stellar age from Uranium decay is at presently limited by incomplete 
knowledge  of oscillator
strengths and production rates of the elements produced in the r-process. 
However, significant progress  can be expected, as theory and observation  shall progress.

\vspace{2cm}
\textbf{Aknowledgment}
~\\

We are extremely  grateful to C. Bonadiman, C. Chiosi, V. Castellani, S. Degl'Innocenti, 
F. Fusi-Pecci, H. Fritsch, G. Ottonello
and  F.L. Villante
  for useful discussions.

%%%%%%%%%%%%%%%%%%%%%%%%%%%%%%%%%%%%%%%%%%%%%

%INDEX%%%%%%%%%%%%%%%%%%%%%%%%%%%%%%%%%%%%%%%%%%%%%%%%%%%%%%%%%%%%%%%
% Please check with the editor of your book whether he plans to
% include a "mutual" subject index - if so, please code your entries
% in the standard syntax. For your own purposes you may print your
% "personal" index by using the following commands:
%
%\clearpage
%\addcontentsline{toc}{section}{Index}
%\flushbottom
%\printindex
%%%%%%%%%%%%%%%%%%%%%%%%%%%%%%%%%%%%%%%%%%%%%%%%%%%%%%%%%%%%%%%%%%%%%

\end{document}